# Surface plasmons of metallic surfaces perforated by nanoholes surface

P. Lalanne, J.C. Rodier and J.P. Hugonin

Laboratoire Charles Fabry de l'Institut d'Optique, Centre National de la Recherche Scientifique, F-91403 Orsay cedex, France

Abstract

Recent works dealt with the optical transmission on arrays of subwavelength holes perforated in a thick metallic film. We have performed simulations which quantitatively agree with experimental results and which unambiguously evidence that the extraordinary transmission is due to the excitation of a surface-plasmon-polariton (SPP) mode on the metallic film interfaces. We identify this SPP mode and show that its near-field possesses a hybrid character, gathering collective and localised effects which are both essential for the transmission.

Keywords : subwavelength aperture, surface plasmons, enhanced transmission

The observation of extraordinary light transmission through metallic films perforated by nanohole arrays [1] has enabled many experimental and theoretical works on the transmission of electromagnetic waves through nano-apertures. Arrays of nanoholes as well as isolated or dressed single holes [2] have been studied in the THz and optics regimes [3] for different aperture shapes [4-5-6], for metallic, dielectric and perfectly-conducting films [7] and for different film thicknesses ranging from $\lambda/2$ to $\lambda/100$. Most of these studies have been devoted to the understanding of the mechanisms responsible for the transmission. This transmission is largely attributed to the excitation of surface plasmon polaritons (SPP) in the literature [8-9-10-11-12-13-14]. These authors assume that the fields associated with a SPP on the top or bottom grating interfaces enhance the evanescent field within the hole and thus contribute to funnel the incident plane wave inside the holes. This currently widely-held view has not been straightforwardly established. The first theoretical studies have dealt with the transmission through slit arrays. For this geometry, which is only apparently related to the nanohole-array geometry used in [1], the first studies [15-16-8-17-18] have concluded that the SPP resonance enhances the transmission. Others have argued that SPP play no casual roles [19] and even a negative role [20] in the transmission enhancement. However, the physics pertaining to slits is fundamentally different to that of circular apertures, and thus the slit-array geometry cannot be used to analyse hole-array geometry, as will be discussed hereafter. Recently, it has been convincingly argued [21] that the extraordinary transmission, which is renowned to provide enhancement factor of order 1000, in fact provides an enhancement factor less than 7 when comparing the transmission properties of a hole in an array to those of an identical single, isolated hole fabricated on the same film. Further the authors in [21] have criticized the SPP-model interpretation and have provided another model based on the interference of evanescent fields generated by the nanoholes. Thus after seven years of intense research and despite a convergence of views, no global consensus exists. In this paper, we identify the SPP mode responsible for the extraordinary transmission through rigorous numerical calculation. We further discuss its nature and we show for the first time that its near-field possesses a hybrid character, gathering collective effects involving neighbourhood hole periodicity and localised effects relying on the fine structure of the aperture geometry. Both characters are essential in the transmission.

We first consider the light transmission through a silver film deposited on a glass substrate and perforated by a nanohole array. The geometry parameters are exactly those of the samples characterized in [13] : periodicity a = 420 nm, circular-hole diameter d = 250 nm and film thickness h = 180 nm. In a linear grey scale, Fig. 1a shows the calculated (0,0)-order transmission as a function of frequency and in-plane wave vector. We have used the dielectric constant of silver tabulated in [22] for the calculation. The results, obtained with a vectorial Fourier-modal method [23] for p-polarization, are in excellent agreement with the experimental data, see Fig. 2a in [13], showing the existence of two-branches in the ($\omega$, $k_{//}$) diagram. Similar calculation have been performed for a thicker self-supported membrane in air with square holes. The grating parameters are h = 500 nm, a = 750 nm and the hole size is d = 280 nm (area d x d). The results are displayed in Fig. 1b. Additionally, the thin white curves in Fig. 1a (resp. 1b) represent the SPP dispersion curves of a flat silver-glass (resp. air) interface. By noting the overall similarity between the high-transmission branches of the ($\omega$, $k_{//}$) diagram and the SPP dispersion curves of flat interfaces, the authors in [13] have interpreted these branches as the signature of the coupling between incident light and SPP modes supported by the perforated interface. Although convincing, the interpretation remains qualitative since the SPP dispersion curves of the flat interface are largely offset by ≈10% with the high-transmission branches. In fact as stressed in [13], the SPP dispersion curves exactly coincide with the low-transmittance (or high-reflectance) branches rather than with the high-transmission branches. This is exemplified for different polarisations by Figs. 2a-2b and 3b in [13] and by Fig. 1 in this work.

Indeed, to understand the relative locations of the low- and high-transmission branches, one has to rely on rigorous calculations of the SPP modes supported by the grating interfaces. Referring to the inset of Fig. 2, we consider an interface between only two regions labelled I and II. Region I corresponds to a uniform dielectric (vacuum or glass with refractive index $n_I$) half-space and region II to a silver half-space perforated by semi-infinite air holes. Briefly, the SPP dispersion relation of the interface is calculated as follows. For a given parallel momentum, $\mathbf{k}_{//} = (k_x\mathbf{x} + k_y\mathbf{y})$, the longitudinal electric field in region I, $\Phi_I(\mathbf{r},z) = (E_x, E_y)$ can be expanded in a plane wave basis (Rayleigh expansion) of the form

$$\Phi_I(\mathbf{r},z) = \Sigma_{m,n}\, \mathbf{a}_{I,m,n} \exp[i\, k_{m,n}\, z]\, \exp[i\,(\mathbf{k}_{//}+\mathbf{K}_{m,n})\,\mathbf{r}] + \Sigma_{m,n}\, \mathbf{b}_{I,m,n} \exp[-i\, k_{mn}\, z]\, \exp[i\,(\mathbf{k}_{//}+\mathbf{K}_{m,n})\,\mathbf{r}]. \quad (1)$$

In Eq. (1), $i^2 = -1$, $\mathbf{K}_{m,n} = 2\pi(m\mathbf{x} + n\mathbf{y})/a$, $(k_{m,n})^2 + (\mathbf{k}_{//}+\mathbf{K}_{m,n})^2 = (k_0 n_I)^2$, and $\mathrm{Re}(k_{m,n}) > 0$, $\mathrm{Im}(k_{m,n}) \geq 0$. The vectors $\mathbf{x}$, $\mathbf{y}$ and $\mathbf{z}$ are the unit vectors of the Cartesian coordinate system. The $\mathbf{a}_{I,m,n}$ (resp. $\mathbf{b}_{I,m,n}$) are 2x1 vectors related to outgoing (resp. incoming) planes waves. Similarly in region II, the electromagnetic field $\Phi_{II}(\mathbf{r},z)$ is conveniently written in a Bloch-mode basis as

$$\Phi_{II}(\mathbf{r},z) = \Sigma_p\, a_{II,p}\, \mathbf{F}_p(\mathbf{r})\, \exp[i\, k_p\, z] + \Sigma_p\, b_{II,p}\, \mathbf{F}_p(\mathbf{r})\, \exp[-i\, k_p\, z], \quad (2)$$

with $\mathrm{Im}(k_p) \geq 0$. In Eqs. 1, the summation includes a finite set of propagative planes waves and an infinite set of evanescent ones. In Eqs. 2, the Bloch modes $\mathbf{F}_p$ are pseudo-periodic functions. They are all evanescent for the subwavelength holes considered in this work and are calculated as eigenvectors of a propagation operator [23]. The associated propagation constants $k_p$'s are the corresponding eigenvalues. Similar expansions exist for the transverse components of the magnetic fields. By matching all tangential electromagnetic-field components at the z = 0 interface, one obtains the scattering matrix **S** which links the outgoing mode amplitudes $a_{I,m,n}$, $a_{II,p}$, to the ingoing ones $b_{I,m,n}$, $b_{II,p}$. We emphasize that the electromagnetic response of the isolated interface is rigorously modelized since all evanescent waves in regions I and II are included in the computation. Numerical errors, which are kept at a negligible level, result only from matrix truncation in the Fourier basis. From the knowledge of the **S** matrix, the SPP dispersion relation is obtained by allowing complex values for the frequency and namely, by searching the complex poles $\tilde{\omega}$ of **S** for fixed real values of $\mathbf{k}_{//}$. Although it shares many technical aspects with that in [8,24], the approach developed hereafter largely differs conceptually as will be discussed in the following.

The solid black curves in Figs. 1a-1b represent the calculated SPP dispersion relation, $\mathrm{Re}(\tilde{\omega})$ as a function of $k_x$. For the square hole geometry, see Fig. 1b, the predicted SPP mode positions *exactly coincide* with the high transmission branches of the spectrum. For the circular hole geometry, an overall *weak deviation* exists. To explain for the weak deviation which is especially visible for small values of $k_x$, we refer to a simplified Fabry-Perot model which has been previously used in [9]. We denote by $\tau_1(k_x,\omega)$ the coupling coefficient between the incident plane wave and the fundamental Bloch mode $\mathbf{F}_1$, by $\tau_2(k_x,\omega)$ the coupling coefficient between $\mathbf{F}_1$ and the transmitted zero-order plane wave, and by $\rho_1(k_x,\omega)$ and $\rho_2(k_x,\omega)$ the reflection coefficients of $\mathbf{F}_1$ at the top and bottom interfaces, respectively. $\mathbf{F}_1$ will be accurately defined hereafter. Under the assumption that the energy flow in the grating region is mediated only via the fundamental Bloch mode $\mathbf{F}_1$, the grating transmission T can be written

$$T = \left| \frac{\tau_1 \tau_2 \exp(ik_0 n_{eff} h)}{1 - \rho_1 \rho_2 \exp(2ik_0 n_{eff} h)} \right|^2, \tag{3}$$

where $n_{eff} = k_1/k_0$ is the effective index of $\mathbf{F}_1$. It is a complex number whose imaginary part mainly represents the evanescent character of $\mathbf{F}_1$. Note that for the silver-membrane-in-air geometry of Fig. 1b, $\rho_1 = \rho_2$ and $\tau_1 = \tau_2$, since the top and bottom interfaces are identical. As we did previously for slit geometries [12], we have rigorously calculated the coefficients $\rho_p$ and $\tau_p$, p = 1,2 for the two geometries of Fig. 1. The transmission predicted by the Fabry-Perot expression of Eq. (3) has been compared with rigorous-calculation data. A quantitative agreement - deviation bellow 0.02 have been obtained for $\mathbf{k}_{//} = 0$ - showing that the approximate expression can be used with confidence for analysing the transmission properties. $\rho_p$ and $\tau_p$ possess the same pole $\tilde{\omega}_p$, which indeed corresponds to the SPP mode of the single interface p, and can take values largely in excess of unity [9] for $\omega = \text{Re}(\tilde{\omega}_p)$. As shown by elementary manipulation of Eq. (3), it is important to note that the pole of T in Eq. 3, which are the same poles as those calculated in [8,24] for the scattering matrix associated to the whole diffraction geometry, are in general not equal to the poles of $\rho_p$ and $\tau_p$. The poles of T result from a combination of SPP interface resonance and of vertical Fabry-Perot resonance due to the recirculation of the evanescent mode $\mathbf{F}_1$ between the two *coupled* interfaces. This explains why the high-intensity branches for the transmission of the silver-film-on-glass geometry, Fig. 1a, do not exactly coincide with the SPP mode dispersion. For the silver-membrane-in-air geometry, Fig. 1b, the analysis is rendered much simpler because the membrane thickness is larger. For this geometry, $\rho_1\rho_2 \exp(2ik_0 n_{eff} h) \ll 1$, and $T \approx |\tau_1 \tau_2 \exp(ik_0 n_{eff} h)|^2$. The poles of T and $\tau_p$ are identical ; this is the reason why the high-intensity branches of the transmission exactly coincide with the SPP mode dispersion in Fig. 1b. As a whole, the quantitative agreement in Fig. 1 between the high-intensity branches of the transmission spectrum and the SPP mode dispersion of the perforated dielectric-metal interface is an important result of this work. It unambiguously attributes the extraordinary transmission to the excitation of the specific SPP modes supported by the grating interfaces.

The perfect coincidence between the low transmittance branches of the ($\omega$, $k_{//}$) diagram and the SPP dispersion curves of flat interfaces is not *fortuitous*. It is even a general property of these metallic films, which has been previously explained through a perturbative model as a non-resonant surface plasmon effect [12] : whenever the momentum of the incident wave matches the SP momentum of a flat interface modulo the grating vector, the incident light does not "see" the array of nanoholes and the perforated interface thus behaves as a flat one. In more mathematical terms, this property results in a nearly real zero for the coefficients $\tau_1$ or $\tau_2$. Consequently, the transmission is very weak ($\approx 10^{-6}$) over a broad spectral interval centred around the flat-interface SP wavelength $\lambda_{sp}$ - T varies as $T \approx (\lambda_{sp} - \lambda)^4$ for membranes in air [12]- and the (0,0)-order reflected intensity is nearly equal to that of a flat interface. For other geometries which support a *propagative* Bloch mode $\mathbf{F}_1$, like films perforated by subwavelength slit arrays and illuminated under transverse-magnetic polarization, the Fabry-Perot model remains valid for thick enough metallic films [12,19], but shows up a completely different physics. The effective index of $\mathbf{F}_1$ becomes nearly real and because of energy conservation, $\tau_p$ or $\rho_p$ are bounded : $|\tau_p|^2 < 1$, $|\rho_p|^2 < 1$, with $|\tau_p|^2 + |\rho_p|^2 \approx 1$. Although the coefficients $\tau_p$ or $\rho_p$ possess a complex pole which can be associated to the SPP mode of the single interface [12], this pole is only marginally related to the high transmission branches. In fact, these branches are related to the poles of the scattering matrix associated to the whole diffraction geometry, or equivalently within the Fabry-Perot model to the zeros of the denominator in Eq. (3). A high transmission occurs for real frequencies fulfilling the Fabry-Perot-resonance condition $2k_0\, \text{Real}(n_{eff})\, h + \arg(\rho_1\rho_2) = 2m\pi$, with m being an integer. Thus the existence of SPP modes which exalt the field at the grating interfaces is not required for high transmission. In general, for such geometries, working at a frequency close to the SPP resonance lowers the transmission, because of additional interface ohmic losses and because of the close frequency proximity with the low-transmission branches, i.e. with the zeros of $\tau_p$. More details can be found in [12,20].

Let us now consider the near-field properties of the SPP mode of the single interface. Because of symmetry, only the low energy SPP mode is excited at normal incidence. The modulus of its dominant magnetic-field component $H_y$ calculated at the dielectric-silver interface (plane z = 0), is shown in Fig. 3a. The $H_y$ field is mainly formed by a background pattern with a symmetric standing wave which corresponds to the interference of two counter-propagating SPP with parallel wave vectors $\pm 2\pi/a$, see the similarity with the white curves representing $|\cos(2\pi x/a)|$. On the upper and lower sides of the aperture entrance, $H_y$ also encompasses a localized hot spot. This hot spot is a consequence of the existence of a strong $H_z$ component in the SPP mode of the perforated dielectric-silver interface, as shown in Fig. 3b. This $H_z$ field strongly looks like that of the fundamental Bloch mode $\mathbf{F}_1$ of the nanohole array, see Fig. 3c. By fundamental mode, we do not intend to mean the less attenuated mode but rather the evanescent mode with a propagation constant $k_1'$ which would be equal to $\sqrt{k_0^2 - (\pi/d)^2}$ if silver would be considered as a perfect metal. Thus, the near-field SPP mode of the perforated dielectric-silver interface possesses a *hybrid* character which results from the combination of a classical delocalised mode propagating along the interface with a localized field originating from the fundamental Bloch mode of the nanohole array. This localization character of the SPP mode well explains the recent observation [4] of a strong dependence of the enhanced transmission with the subwavelength aperture geometry. It also strongly affects the gap width between the high and low energy branches of the SPP modes for $k_{//} = 0$. For the circular geometry which possesses a rather large relative aperture area, $\pi d^2/(4a^2) = 28\%$, the relative gap width is only $\approx 0.9\%$. Although its relative aperture area is smaller, $d^2/(a^2) = 14\%$, the square geometry offers a much larger relative gap width of $\approx 4\%$.

Clearly the nanohole geometry impacts the transmission properties. Periodicity also plays a salient role in the transmission at least for relatively thick films. In Fig. 4, we compare the transmission enhancement of a silver membrane in air perforated by an array of square nanoholes (Fig. 4a) with that of the same membrane perforated by a single nanohole (Fig. 4b). The results hold for an illumination by a plane wave of constant intensity $I_0$ under normal incidence. The power T transmitted through the array geometry, calculated with the method of Ref. 23, represents the (0,0)-order contribution. In the spectral range of interest, only the (0,0)-order is propagating, thus T is the total power transmitted. The calculation of the single-hole geometry has been performed with a rigorous Fourier modal method by using an artificial periodization and by incorporating Perfectly-Matched-Layers between adjacent periods [25]. The power T transmitted though this non-periodic geometry is obtained as the flux of the Poynting vector z-component in a transversal plane (z = 0) just below the aperture. The transmission enhancement $T/P_0$ is defined as the ratio between the power T transmitted through the membrane and the power $P_0 = I_0 d^2$ from the incident beam impinging on each aperture. Thus $T/P_0$ represents the per-hole relative transmittance. Two general trends are observed. For a small membrane thickness, the maximum relative transmission of the periodic structure is 3-time larger than that of the single hole geometry (for $\lambda$ = 0.86 μm). On average over the spectral range considered, it is even weaker. These results are consistent with recent experiments [21] which compare the transmission of an array of hole with that of a real, identical and isolated hole in the same film and which evidence that placing a hole in an array leads to a marginal enhancement of the transmission. The situation is radically different for a large thickness. For instance, the relative transmittance of the periodic array is 35-time larger than that of the single hole geometry for $\lambda$ = 0.81 μm and h = 0.5 μm. For thicker membranes, an enhancement of order 100 is predicted by other computational results not reported here. In our opinion, the term "extraordinary" makes sense in this context. We are not aware of any measurements confirming this prediction for thick films, but this would be interesting to confirm.

To summarize, we have provided a rigorous electromagnetic study of the SPP modes supported by dielectric-metallic interfaces perforated by nanohole arrays. These SPP modes have been shown to be responsible for the so-called extraordinary transmission though metallic films with holes in them. We have shown that the SPP near-field possesses a hybrid character, gathering collective and localised effects which are both essential for the transmission.

The authors acknowledge financial supports from the European Network of Excellence NEMO and from the project DESIG under the French programme NanoSciences 2003.

# Figures

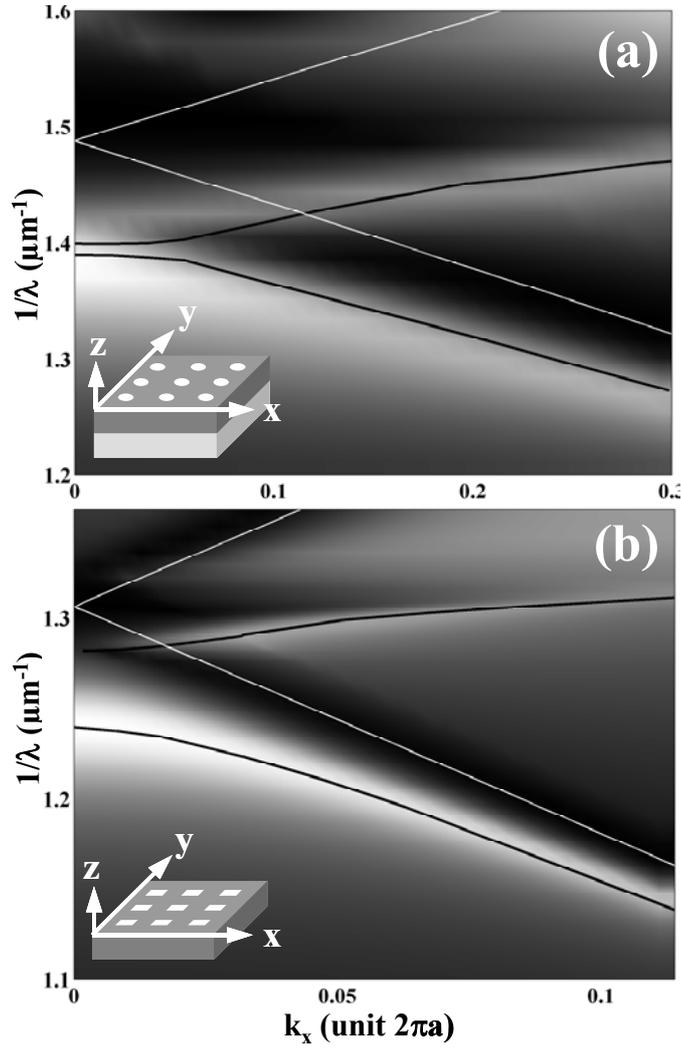

**Fig. 1**. Grey-scale images showing the (0,0)-order far-field transmittance of two silver gratings as a function of the frequency and the in-plane wave vector $k_x$. **(a)** Circular hole geometry for a silver-film on glass. Experimental results can be found in Fig. 2a of Ref. [13]. The scale is linear and the maximum transmittance is 32%. **(b)** Square hole geometry for a silver-membrane in air. The maximum transmittance is 50%.

Superimposed black curves : corresponding SPP dispersion-relation of a perforated dielectric-silver interface, shown in Fig. 2, glass-silver perforated by circular air-holes for (a) and air-silver perforated by square air holes for (b). The white curves correspond to the SPP dispersion-relation of a flat interface, glass-silver for (a) and air-silver for (b).

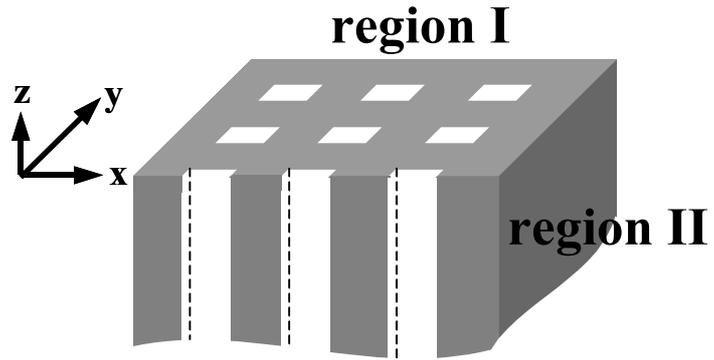

**Fig. 2**. Perforated dielectric-silver interface. Region I corresponds to a uniform dielectric (vacuum or glass with refractive index $n_I$) half-space and region II to a silver half-space perforated by semi-infinite air holes.

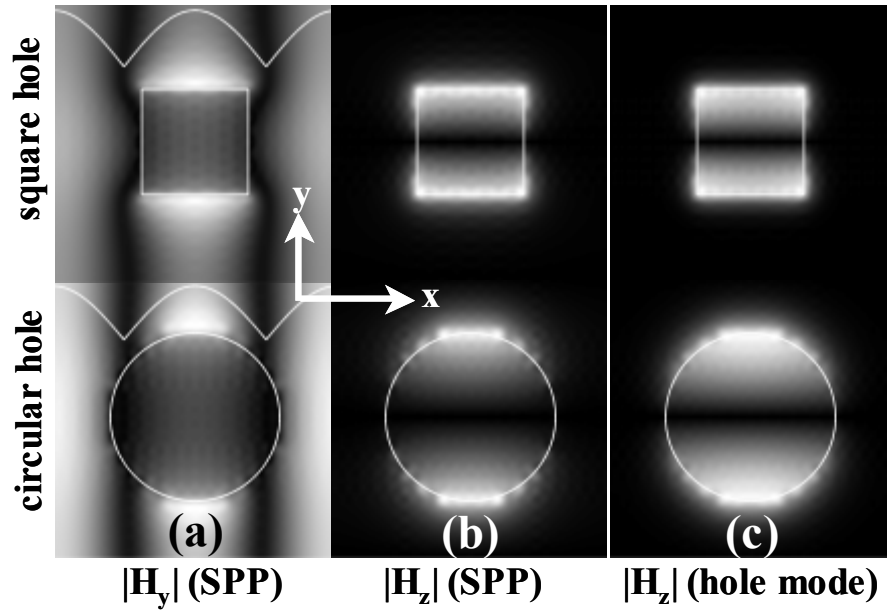

**Fig. 3**. Localization versus delocalisation properties of the low energy SPP mode for $k_{//} = 0$. The near-fields are plotted on the interface ($z = 0$) and every image covers a single period. They are obtained for a (0,0) Rayleigh plane-wave polarized along the x-direction, $\mathbf{a}_{I,0,0}\cdot\mathbf{y} = 0$ in Eq. (1). **(a)** $|H_y|$ of the SPP mode **(b)** $|H_z|$ of the SPP mode. **(c)** $|H_z|$ for the fundamental Bloch mode of the nanoholes. The superimposed white curve in (a) corresponds to $|\cos(2\pi x/a)|$.

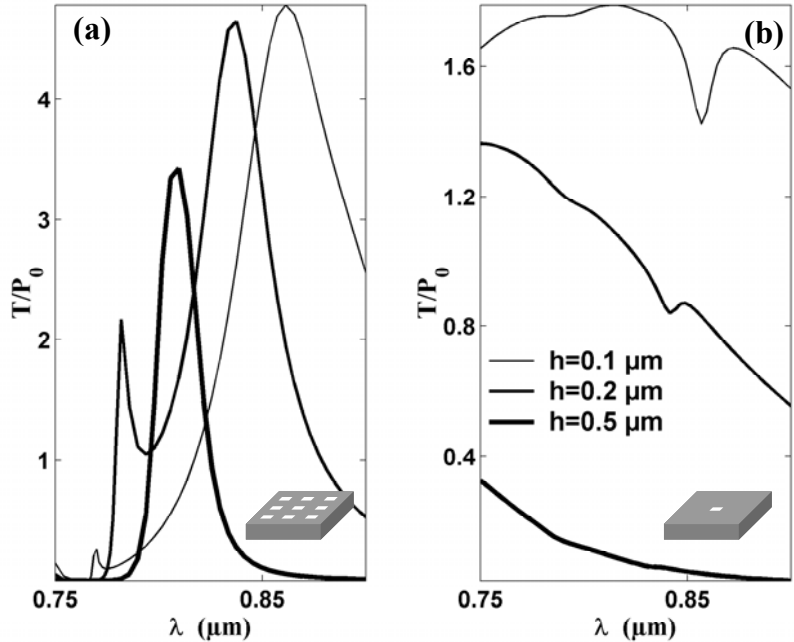

**Fig. 4**. Per-hole total transmission $T/P_0$ for a periodic array of square holes **(a)** and for a single square aperture **(b)**. The results are obtained for silver membranes in air and for several thickness values h = 0.1, 0.2 and 0.5 μm.

# References

[1] T. W. Ebbesen, H.J. Lezec, H.F. Ghaemi, T. Thio and P.A. Wolff, Nature (London) **381**, 667 (1998).

[2] T. Thio, K.M. Pellerin, R.A. Linke, T.W. Ebbesen, and H.J. Lezec, Opt. Lett. **26**, 1972-74 (2001).

[3] J. Gmez-Rivas, C. Schotsch, P. Haring Bolivar, and H. Kurz, Phys. Rev. B **68**, 201306(R) (2003).